\documentclass[useAMS,usenatbib,usegraphicx]{mn2e}
\usepackage{amsmath}
\usepackage{amssymb}
\usepackage{graphicx}
\usepackage{latexsym}
\usepackage{aas_macros}
\usepackage{natbib}
\usepackage[squaren]{SIunits}
\usepackage{array}
\usepackage{tabularx}

\newcommand{\scinot}[2]{\ensuremath{#1 \times 10^{#2}}}

\newcommand{\Tkep}[1]{\ensuremath{T_{\text{kep}}^{#1}}}
\newcommand{\vkep}[1]{\ensuremath{v_{\text{kep}}^{#1}}}

\newcommand{\Mstar}[1]{\ensuremath{m_{*}^{#1}}}
\newcommand{\Mplanet}[1]{\ensuremath{m_{\text{plan}}^{#1}}}
\newcommand{\Mpart}[1]{\ensuremath{m_{\text{part}}^{#1}}}

\newcommand{\paren}[1]{\left ( #1 \right )}
\newcommand{\parenfrac}[2]{\paren{\frac{#1}{#2}}}

\addunit{\Mearth}{\ensuremath{\mathrm{M}_{\earth}}}
\addunit{\AU}{au}

\addunit{\cm}{\centi\metre}
\addunit{\cmpersecnp}{\cm \usk \reciprocal \second}
\addunit{\yyear}{yr}

\newcommand{\mySun}{\odot}

\addunit{\Msol}{\ensuremath{\mathrm{M}_{\mySun}}}

\title{Pumping of a Planetesimal Disc by a Rapidly Migrating Planet}
\author[R.G.~Edgar \& P.Artymowicz]{Richard~Edgar$^1$\thanks{Email: rge21@astro.su.se} and Pawel Artymowicz$^1$ \\
$^1$Stockholm Observatory} 
\date{\today}

\pagerange{\pageref{firstpage}--\pageref{lastpage}} \pubyear{2004}

\begin{document}

\label{firstpage}

\maketitle

\begin{abstract}
We examine the effect of a rapidly migrating protoplanet on a ring of planetesimals.
The eccentricities of the planetesimals are usually increased by $\Delta e \in (0.01, 0.1)$, with the exact increase being proportional to the protoplanet's mass, and inversely proportional to its migration rate.
The eccentricity distribution is also substantially changed from a Rayleigh distribution.
We discuss the possible implications for further planet formation, and suggest that the rapid passage of a protoplanet may not prevent the planetesimal disc from forming further planets.
\end{abstract}

\begin{keywords}
Solar system: formation -- planets and satellites: formation
\end{keywords}


\section{Introduction}

A planet in orbit around a star will disturb the orbits of test particles.
This situation, known as the `restricted three body problem' has been studied for centuries (see, e.g. \citet{1999ssd..book.....M}).
Particles close to resonant locations have their orbital elements changed due to repeated interactions with the planet.
These changes may be computed using various analytic techniques (\emph{ibid}), and the analysis can be extended to slowly migrating planets.
Recently, a rapid migration mode for protoplanets has been discovered \citep{2003ApJ...588..494M}.
The planet can have its semi-major axis halved in only a hundred orbits or so, which is far too fast for the effect on planetesimal orbits to be calculated analytically.

In this paper, we attempt to quantify the expected eccentricity increase in a planetesimal ring, due to a rapidly migrating protoplanet.
A simple theoretical calculation is described in section~\ref{sec:theory}.
We describe a numerical model in section~\ref{sec:model} and the results obtained from it in section~\ref{sec:results}.
The possible implications for the formation of the Solar System are discussed in section~\ref{sec:implications}.
We summarise our findings in section~\ref{sec:conclude}.

\section{Theoretical Predictions of Planetesimal Excitation}
\label{sec:theory}

The orbits of the planetesimals can be be changed in two ways:
\begin{enumerate}
\item Long term interactions with resonances
      (mean motion, secular and co-rotation)
\item Close encounters with the planet
\end{enumerate}
Interactions between test particles and resonances have been studied for several centuries.
However, for a rapidly migrating planet, these resonances are much less important, since they will be sweeping through the disc with the migrating planet.
Hence, the resonance is unlikely to remain in the vicinity of a particle long enough to make significant changes.
This is supported by the work of \citet{1999Icar..139..350T}, who found that a slow migration rate led to `shepherding,' (the planet gently brings the planetesimals with it), while fast migration leads to `predatory' behaviour (the planet simply ploughs through the particle disc) - see also the work of \citet{1995ApJ...440L..25W}.
In our calculations, migration is even faster than the `rapid' migration of \citet{1999Icar..139..350T}.

\citet{1990A&A...227..619H} found that the expected change in eccentricity of a single particle per collision was given by
\begin{equation}
\langle \Delta e^{2} \rangle
=
81 R_1^2 \frac{h^6}{b^4}
\label{eq:HNdeltaEcc}
\end{equation}
where $R_1 = 0.747$, $h=(\Mplanet{}/3\Mstar{})^{1/3}$ is the Hill parameter, and $b$ is the impact parameter of the collision, in units of the semi-major axis.
This is equation~38 of their paper, rewritten with conventional (rather than normalised) eccentricities.
It is derived on the assumption that the stellar mass ($\Mstar{}$) dominates, that the eccentricities and inclinations are small, and that the impact parameter is large enough to ensure that the particles can't enter their mutual Hill sphere.
\citeauthor{1990A&A...227..619H} give another, similar, formula for the expected increase in inclination, but the coefficient is much smaller and we shall neglect inclination here (note also that the eccentricity increase is independent of the inclination).
Equation~\ref{eq:HNdeltaEcc} (with a different coefficient) may also be derived following the impulse approximation of \citet{1979MNRAS.186..799L}.
Note the strong dependence on the impact parameter (curiously reminiscent of Rutherford scattering, although the link is not direct).
This suggests that it is the closest encounter between the planet and planetesimal which will be the most important.

But what is the closest encounter we should expect?
Ever closer encounters will give larger changes (although note that equation~\ref{eq:HNdeltaEcc} eventually ceases to be valid), but encounters with small impact parameters are rare.
Both the planet and planetesimals are on near-circular orbits, so an impact parameter of zero implies that the two bodies are on the same orbit.
Such particles would have a (near) infinite synodic period.
Consider a particle at semi-major axis $a_0$, and the planet migrating from $a_0 + \Delta a$ to $a_0 - \Delta a$.
We can expect an interaction with $b \approx \Delta a / a_0$ if the time for the planet to migrate between the two limits ($2 \Delta a / \dot{a}$) is equal to the synodic period for orbits at $a_0$ and $a_0 + \Delta a$.
Formally, this neglects the orbital inclination, but we expect the effect to be small in a disc.
We have
\begin{eqnarray}
T_{\text{syn}}( a_0, \Delta a) & = & T_{\text{cross}} \\
\frac{2 a}{3 \Delta a} \Tkep{} & = & \frac{2 \Delta a}{\dot{a}} \\
\parenfrac{\Delta a}{a_0}^{2} & = & \frac{\dot{a}}{3 a_0} \Tkep{}(a_0)
\end{eqnarray}
where $\vkep{}$ is the circular Keplerian velocity, and $\Tkep{}$ is the Keplerian orbital period.
Substituting into equation~\ref{eq:HNdeltaEcc}, we find
\begin{equation}
\langle \Delta e^2 \rangle^{\frac{1}{2}}
=
\frac{27 R_1 h^3}{\Tkep{}(a_0)} \parenfrac{a_0}{\dot{a}}
\label{eq:SimplePredictDeltaEcc}
\end{equation}

For a \unit{50}{\Mearth} planet in orbit around the sun and migrating at \unit{10^4}{\cmpersecnp} through a ring close to \unit{3.5}{\AU},\footnote{Note that $\unit{10^4}{\cmpersecnp} \approx \unit{0.2}{\AU\usk\reciprocal\yyear}$, which is not unreasonable for the rapid migration discussed by \citet{2003ApJ...588..494M}} equation~\ref{eq:SimplePredictDeltaEcc} suggests $\langle \Delta e^2 \rangle^{\frac{1}{2}} \approx 0.03$.
Several approximations were made in deriving equation~\ref{eq:SimplePredictDeltaEcc}.
It assumes a single close encounter, and the estimated timescale for this (based on the synodic period) is rather imprecise.
The numerical factor could easily be incorrect.
At this point, numerical simulations become useful in determining the evolution.

\section{Model}
\label{sec:model}

To simplify the system as much as possible, we considered
the interaction of a migrating planet with a narrow
ring of test particles.

The bodies in our simulation belonged to one of three types:
\begin{enumerate}
\item The central star, $\Mstar{}$
\item The migrating planet, $\Mplanet{}$
\item The test particles (planetesimals), $\Mpart{}$
\end{enumerate}
We evolved our system using Newtonian gravity,
plus a torque to migrate the planet.
To avoid the computational expense of a full
n-body calculation, the test particles did not
interact with each other.\footnote{For the short
timescales we are studying this is reasonable,
since the close proximity of the (relatively) massive planet
will have a far greater effect}
However, the gravitational interactions between
the star and planet, the star and the particles,
and the planet and the particles, were all
fully computed.

We neglected the effect of gas damping, since we were interested in the effects of \emph{rapid} migration.
This is reasonable, since our largest migration timescales were still far shorter than the damping timescales.
Using the notation of \citet{1999Icar..139..350T}, we typically had $\bar{\tau}_{\text{mig}} < 1$ while $\bar{\tau}_{\text{gas}} \sim 1000$, even with enhanced gas density.
This holds even if the gas damping timescale is assumed to be due to wave excitation, as will be the case for the more massive planetesimals \citep{1993ApJ...419..166A}.
Hence we did not expect the gas damping to be sufficient to make our migrating protoplanets `shepherd' the planetesimals.


\subsection{Orbital Migration}

Orbital migration of the planet is imposed by
applying a torque to the orbit of the star and
planet.
In the interests of simplicity, we apply a torque
chosen to give a
constant migration rate, $\dot{a}$.
The required torque may be computed using
Kepler's Laws:
\begin{equation}
\frac{\dot{a}}{a} = 2 \frac{G}{L}
\end{equation}
where $L=\mu a^2 \Omega$ is the orbital angular momentum, and $G$ is the
required torque.
The torque was applied as an extra force
to the motion of the star and planet.
Migration was halted once the planet's
semi-major axis dropped below a preset value.


\subsection{Intial Conditions and Integration}

The star (\unit{1}{\Msol}) and planet were positioned in a
circular orbit about their centre of
mass.
The initial orbital separation was usually \unit{6}{\AU}, and migration stopped at \unit{0.5}{\AU}.
We distributed the test particles in
circular orbits around the centre of mass
of the star-planet system.
The particles were uniformly distributed
in azimuth and radius (usually \unit{3-4}{\AU}).
Finally, we added small perturbations to
their circular motion, following the
prescription of \citet{2000Icar..143...28S},
to give a Rayleigh distribution of $e$
and $i$ (\emph{ibid}).
We tested a variety of planetary masses and migration rates.

We integrated the equations of motion using
a Runge-Kutta integrator with adaptive
step-size.
To avoid catastrophically small timesteps,
test particles which strayed too close
to the star or planet were eliminated.
Similarly, particles which reached large
radii were removed.
At the end of each timestep, we computed the values
of $e$, $i$ and $a$ for each particle
using the Laplace-Runge-Lenz
formalism.

\section{Results}
\label{sec:results}

\subsection{General Behaviour}

When the planet was far from the planetesimal ring, waves in $(e,a)$ space were observed propagating through the ring.
These were caused by resonances sweeping through the ring - if the planet did not migrate, then sharp peaks in the eccentricity were observed, corresponding to resonant orbits.
However, the increase in eccentricity prior to the planet's encounter with the ring was fairly low.
As the planet ploughed through the ring, the planetesimals were catapulted up lines of constant Jacobi energy, $E_{\text{J}}$, given by \citep{1977PASJ...29..163H}:
\begin{equation}
E_{\text{J}} = \frac{1}{2}\paren{e^2 + i^2} - \frac{3}{8}(a'-1)^2 + \frac{9}{2}h^2
\end{equation}
where $a'$ is the particle semi-major axis in units of the planet's semi-major axis (this energy is expressed in scaled units).
The encounters with the planet were generally at a distance further than the $L_2$ point (this is consistent with the assumptions made in deriving equation~\ref{eq:HNdeltaEcc}).

\subsection{Numerical Results}

Several runs were performed, with the parameters and results summarised in Table~\ref{tbl:NumResults}.
For all these runs, the ring of particles lay between 3 and \unit{4}{\AU} initially.
In this table $\xi^2 = e^2 + i^2$, the conventional measure of deviation from circular, coplanar orbits.
In practice, $\xi$ was dominated by eccentricity, not inclination (cf Section~\ref{sec:theory}).
We used a Kolmogorov--Smirnov test to investigate if the final distribution of $\xi$ was still drawn from a Rayleigh distribution.
In every case, this hypothesis was strongly rejected.\footnote{We also checked that the KS test \emph{did} allow the initial distribution to be Rayleigh}
Qualitatively, we found that there was a tail of high $\xi$ particles.
This made the RMS values for $\xi$ somewhat changeable, so we quote the median value as well.
For the remainder of this discussion, references to the final $\xi$ values refer to the median.

\begin{table*}
\caption{Numerical Results. The quantity $\xi$ is defined as $\xi^2 = e^2 + i^2$. Recall that $\unit{10^4}{\cmpersecnp} \approx \unit{0.02}{\AU\usk\reciprocal\yyear}$}
\begin{minipage}{126mm}
\begin{tabular}{lcccccc}
Run  &   $\Mplanet{} / \Mearth$    &  $\dot{a} / \cmpersecnp$ &  Initial $e,i$ RMS  &  Final $\xi$ RMS  & Final $\xi$ median \\
\hline
1    &   100   &  $10^{4}$    &   $10^{-3}$    &   0.0859   &  0.0360 \\
2    &   100   &  $\scinot{5}{3}$ & $10^{-3}$  &  0.1472   &  0.0978 \\
3    &   100   &  $10^{4}$    &   $\scinot{2}{-3}$ & 0.0931 & 0.0358  \\
4    &   100   &  $\scinot{5}{3}$ & $\scinot{2}{-3}$ &  0.1717  & 0.0987  \\
5    &   200   &  $10^{4}$    &  $10^{-3}$ & 0.1549  &  0.0839 \\
6    &   10   &   $10^{4}$    &  $10^{-3}$ & 0.0238  &  0.0039   \\    
7    &   10    &  $\scinot{5}{3}$ & $10^{-3}$ & 0.0612 & 0.0071  \\
8    &   10  &  $\scinot{5}{3}$ &  $10^{-2}$  & 0.0352 & 0.0163 \\
9    &   10  & $\scinot{2}{3}$   &  $10^{-3}$ & 0.0411 & 0.0180 \\
\end{tabular}
\end{minipage}
\label{tbl:NumResults}
\end{table*}

Most of the results of Table~\ref{tbl:NumResults} bear out the \emph{qualitative} behaviour suggested by equation~\ref{eq:SimplePredictDeltaEcc}, but differ quantitatively.
The final median $\xi$ values are close to those predicted (recall that $\xi$ was dominated by $e$ in our simulations).
Increasing the migration timescale or the mass of the migrating planet increases the final value of $\xi$.
However, this doesn't quite happen in the linear manner predicted by equation~\ref{eq:SimplePredictDeltaEcc}.
The detailed values are similar to those predicted, but not identical.

\begin{figure}
\centering
\includegraphics[scale=0.5]{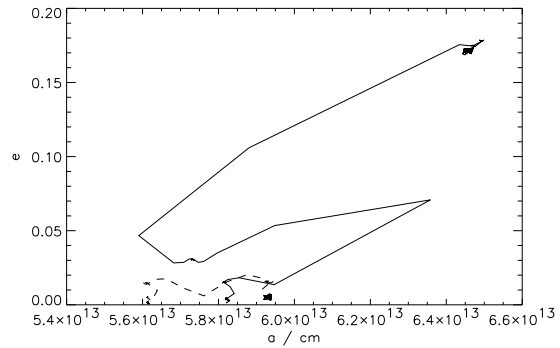}
\caption{Two sample particle trajectories in $e\mbox{--}a$ space}
\label{fig:SampleParticleTrajectory}
\end{figure}

Figure~\ref{fig:SampleParticleTrajectory} shows part of the problem, plotting two sample particle trajectories in $e\mbox{--}a$ space.
These particles have moved around significantly in $e\mbox{--}a$ space, and have not done this just once.
The trajectories are characterised by large jumps (corresponding to those analysed above) interspersed with
periods where the particles are almost stationary.
Sometimes, the eccentricity is pumped up more slowly as well (presumably when the particle happened to be close to one of the resonances of the migrating planet).
This is in sharp contrast to the `single encounter' approximation of Section~\ref{sec:theory}.
Obtaining better predictions of the final $\xi$ value is therefore problematic.
Even if the resonant pumping is neglected, we cannot apply a simple random walk approach.
Firstly, the number of steps (large `jumps') is also random - and not large.
Worse, the steps aren't truely random, but are constrained to have constant Jacobi
energies (although the constant value changes for each step, since the planet will have
migrated in the meantime).

Run~8 behaved slightly differently.
However, even this is not surprising when compared to run~7.
The two runs were identical apart from the initial $\xi$ value, and the \emph{initial} $\xi$ for run~8 was higher than the \emph{final} value for run~7.
It is therefore not surprising that the final $\xi$ value for run~8 is higher than that for run~7 - but notice that the \emph{increase} in $\xi$ is similar for both runs.

\begin{figure}
\centering
\includegraphics[scale=0.4]{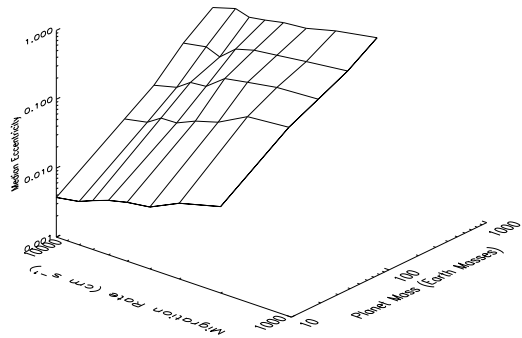}
\caption{Surface of median $\xi$ values following migration}
\label{fig:MedianXiSurface}
\end{figure}

We also ran a grid of 40 models, with $\Mplanet{}$ ranging between 10 and \unit{400}{\Mearth} and $\dot{a}$ in the range $\scinot{2}{3}$ to \unit{10^4}{\cmpersecnp}.
This covers the range expected by \citet{2003ApJ...588..494M} to undergo runaway migration, and a bit more on each end.
All these runs started with average $e$ and $i$ values of $10^{-3}$.
Figure~\ref{fig:MedianXiSurface} plots a surface showing the resultant median $\xi$ values.
Planets more massive than \unit{100}{\Mearth} frequently managed to increase $\xi$ to be greater than 0.1 (and reached 0.4 for a \unit{400}{\Mearth} planet migrating at \unit{\scinot{2}{3}}{\cmpersecnp}).
Although not a perfect match, the behaviour predicted by equation~\ref{eq:SimplePredictDeltaEcc} is seen.

\section{Implications}
\label{sec:implications}

One major problem with the migration of planets in their nascent discs is that protoplanets ``could be too mobile for their own good'' \citep{2000prpl.conf.1135W}.
The rapid migration mode of \citet{2003ApJ...588..494M} makes the protoplanets even more mobile - it seems to be very easy to lose protoplanets into the Sun.
Whilst this is obviously not very `good' for the protoplanets in question, would such events make the formation of the Solar System impossible?
Put another way, ``Is the current Solar System the first Solar System?''

Considering the $Q$ parameter of \citet{1964ApJ...139.1217T} suggests that the disc around the young Sun may have been up to five times more massive than the Minimum Mass Solar Nebula (MMSN) model of \citet{1981PThPS..70...35H} (although \citet{1985prpl.conf.1100H} is likely to be a more accessible reference).
This would give plenty of material for forming several generations of planets.
The question is therefore whether a migrating planet will disrupt the disc sufficiently to prevent further planet formation within the nebula lifetime of a few $\mega\yyear$.

If the migration rate is low enough to permit shepherding (and hence depletion of planetesimals inward of the protoplanet's intial location), \citet{2003ApJ...582L..47A} has shown that it is unlikely that fresh material could diffuse into the inner portions of the disc.
In this case, further planet formation would not be possible.
We have considered a migration rate too high for shepherding, so we must examine whether our planetesimal disc is too hot to allow further planet formation.

In our simulations, $\Delta e$ is typically in the range $(0.01, 0.1)$, dependent on the mass of the protoplanet and its migration rate.
Planetesimals are fairly fragile objects - bodies with $\Mpart{} \approx \unit{10^{22}}{\gram}$ will suffer disruptive collisions if $e \gtrsim 0.01$ \citep{2001Icar..153..416K}.
However, gas damping can recircularise the orbits of such bodies very rapidly - probably less than \unit{10^4}{\yyear} (see figure~1 of \citet{1999Icar..139..350T} and references therein) - especially if the disc is denser than the MMSN.
The damping timescale will reach a peak for $\Mpart{} \approx \unit{10^{25}}{\gram}$, but these bodies require $e \gtrsim 0.1$ before suffering disruptive collisions.
Still larger bodies will require even higher $e$ values before they shatter, but these damp faster due to the excitation of density waves \citep{1993ApJ...419..166A}.
Gravitational focussing will also fall as eccentricities increase, due to the higher relative velocities implied.
This will cause a dramatic fall in the collision rate (cf figure A1 of \citet{1997Icar..128..429W}), so the likelihood of disruptive collisions is reduced.
Finally, in a real planetesimal disc there will be a distribution of sizes.
\citet{1993Icar..106..190W} found that grinding moderately sized planetesimals into rubble helps the larger planetesimals accrete them, so some excitation may even be helpful.

Based on this discussion, it seems that the rapid passage of a protoplanet will not prevent further planet formation.
Coagulation will be suppressed for a while, but this should be short compared to the disc lifetime.
However, the arguments in the preceding paragraph are not rigorous, and a longer term calculation of the `end' states would be required to give a firm answer.
This is particularly true just as the planetesimal disc re-achieves equilibrium.

\section{Conclusions}
\label{sec:conclude}

Combining our results (including some varying $a_0$ not explicitly listed here), we conclude that a rapidly migrating protoplanet will increase the eccentricity of a planetesimal ring by roughly
\begin{equation}
\Delta e \approx \scinot{7}{-4} \parenfrac{\Mplanet{}}{\Mearth} \parenfrac{a_0}{\unit{1}{\AU}}^{-\frac{1}{2}} \parenfrac{\dot{a}}{\unit{10^4}{\cmpersecnp}}^{-1}
\end{equation}
where $a_0$ is the initial semi-major axis of the planetesimal ring and $\dot{a}$ is the protoplanet's migration rate.
During the passage through the ring, the distribution of eccentricities will also be substantially changed from a Rayleigh distribution.
In particular there will be a tail of high eccentricity bodies.

Simple arguments suggest that this effect should not prevent subsequent planet formation by the surviving planetesimals (very few are ejected or accreted).
The resultant eccentricities should not be dangerously high for at least some of the planetesimals, collisions are less likely due to reduced gravitational focussing, and damping timescales are short.
However, problems could arise shortly before the planetesimal disc achieves equilibrium once more, and further simulations (spanning longer timescales and including all gravitational interactions) are needed to assess the effect in detail.


\bibliography{coreaccretion,gravinstability,theory,compute,nbody,migration,misc,reviews}
\bibliographystyle{astron}


\section*{Acknowledgements}

RGE acknowledges financial support provided through the European Community's Human Potential Programme under contract HPRN-CT-2002-00308, PLANETS

Some of the calculations used the resources of High Performance Computing Centre North (HPC2N) and National Supercomputing Centre, Link\"{o}ping.

The KS test was performed using a publically available routine written by W.~Landsman.

The authors would like to acknowledge helpful discussions with Willi Kley, which took place while attending the KITP workshop on planet formation.

\bsp

\label{lastpage}

\end{document}